\documentclass{aa}  
\usepackage{graphicx}
\usepackage{hyperref}
\usepackage{txfonts}
\usepackage{lipsum}
\usepackage{subcaption}   
\usepackage{amsmath}
\usepackage{lscape}             
\usepackage{placeins}           
                                
\begin{document}

\makeatletter
\let\linenumbers\relax
\let\nolinenumbers\relax
\makeatother

   \title{Addressing Exoplanetary Degeneracies Through Geometry}

   \subtitle{Application to Asynchronously Rotating Systems}

%
%
%

   \author{Mradumay Sadh\inst{1}\corrauth{mradumay.sadh@students.mq.edu.au}        
        }

   \institute{School of Mathematical and Physical Sciences and Astrophysics and Space Technologies Research Centre, Macquarie
University, 2109, NSW, Australia
   }


  \abstract{This study augments the geometry-based \textit{InstellCa} code that calculates accurate irradiance values to \textit{InstellCa--2}, which computes the longitude averaged instellation over a given latitude on the planet along with its effective temperature profile. The estimation of instellation is performed considering the planet as a 3D body that asynchronously rotates, causing a varying zenith angle of the star resulting in a diurnal cycle. This serves as a specialised case study to demonstrate how the combination of this relative rotation and extreme proximity to host-stars can affect the estimated effective temperatures of close-in planets. 55 Cancri e is selected specifically to demonstrate the importance of this effect as it has been earlier hypothesised to exhibit asynchronous rotation and also has a debated scientific discourse about the existence of an atmosphere on the planet. The TRAPPIST-1 system, being a multi-planetary system susceptible to mean motion resonances, also serves as an interesting exemplar case to demonstrate these effects. Incorporating the stellar zenith angle to the geometrical analysis allows the possibility of physically feasible Bond albedo values that would otherwise be prohibited if a synchronously rotating bare rock scenario is assumed. Consequently, the degeneracies in the modelling of different atmospheric states of such planets can be reduced if the correct geometry is taken into account.
}


   \maketitle

\section{Introduction}
\label{sec1}

\noindent The vast variety of exoplanetary systems, which have been detected in the previous decades, have helped refine planetary climate models as they collectively exhibit a broad range of orbital and planetary parameters. These discoveries have enabled the demotion of climate models that use Earth-like conditions and assumptions, such as the point-source stellar radiation approximation. Tidally locked and highly irradiated hot-Jupiters, super-Earths, and rocky planets show vastly different possibilities for their atmospheric composition \citep{angelo2017case,hammond2017linking, Parmentier2018,kreidberg2019absence, nguyen2020irradiation,zieba2023no}. New observations and insights have called for better atmospheric and climate models that can be reliably used for the recent high-precision JWST observations \citep{kempton2024transiting}. 

The inclusion of geometry in photometric studies of binary stars \citep{kopal1954photometric,wood1973reflection,wilson1990accuracy, huang2000effect} and thermal phase curve studies of exoplanets \citep{budaj2011reflection,knuth2017exonest, carter2019irradiance, carter2024irradiation, sadh2026importance} is highly pertinent for understanding the potential atmospheric composition of close-in planets. \citet{sadh2026importance} created a general geometric model, tailored specifically for exoplanetary systems with tenuous or absent atmospheres. However, the implications of their model remain in the preliminary \citep{lin2026persistent} stages unless the model is assimilated into general circulation models (GCM). Nonetheless, a natural extension of this 2D irradiance model to 3D longitude-averaged instellation yields interesting and pertinent results.

The debate on the existence of an atmosphere on the planet 55 Cancri e has went on for more than a decade. Multiple hypotheses and models have been proposed to explain the observational thermal phase curves and infrared spectra; ranging from a tenuous atmosphere made of vaporized rock \citep{Zilinskas2022} to a thick atmosphere with clouds \citep{angelo2017case,hammond2017linking}. Consequently, 55 Cancri e has been studied extensively. Recent developments have raised the idea of asynchronous rotation for the planet \citep{patel2024jwst,ferraz2025tidal}. Moreover, improved irradiance models, that account for the close-in geometry of the planet have highlighted that the heat transport models for 55 Cancri e and similar planets should be reconsidered \citep{sadh2026importance}.

The question that this article aims to answer is the following: How can the derived dayside effective temperatures of atmospherically ambiguous planets like 55 Cancri e be explained using a minimalistic but accurate irradiation model? By limiting the scope of our study we bypass the complexities of thermal phase curve analysis such as the observed photometric variability in thermal phase curves \citep{patel2024jwst} of 55 Cancri e, which require advance analysis and precise priors for both stellar and planetary parameters. I extend the 2D planet model proposed in \citet{sadh2026importance} to a 3D model with asynchronous rotation and demonstrate the diurnal temperature profile across the planet. This can also be viewed as increasing the physical degrees of freedom of the previous model. Although the effects of asynchronous rotation on instellation have been addressed in previous exoplanet literature \citep{dobrovolskis2015insolation}, the inclusion of extended-source geometry raises a new aspect of non-zero thermal offset, as will be discussed in this article. These results are supported by the analytical interpretation of the irradiance profiles that use the foundations of the radiative transfer laid out by \citet{sadh2026importance}. Further, I summarise current literature for 55 Cancri e and demonstrate how considering it as an asynchronously rotating planet can exhibit the same observational imprint in the estimated day-side temperature as incorporating atmospheric heat transport effects \citep{hu2024secondary}.  Furthermore, the thermal profiles for the planets of the TRAPPIST-1 system are also calculated. A particularly intriguing and extreme example is TOI-2431 b, which shows the highest offset, among all candidates of the current exoplanet population, due to its extremely small semi-major axis.

\section{Instellation and Effective Temperature Profile}

\noindent Mean motion resonance can perturb spin-orbit synchronisation in tidally locked planets \citep{shakespeare2023day}. Previous studies have highlighted the case of asynchronous rotation for 55 Cancri e \citep{patel2024jwst,ferraz2025tidal}. In particular, they have been cautious not to eliminate the possibility of asynchronous rotation. The non-linear effects of the restricted three body problem \citep{mardling2008resonance} also remain under-explored for the 55 Cancri system, which exhibits rich multi-planetary dynamics \citep{nelson201455,zhao2023measured}.

We assume that the host star, 55 Cancri A, rises and sets periodically on the planet, causing a diurnal variation over an orbit, which is hereafter referred to as \textit{starrise} and \textit{starset}. We also assume the case of zero planetary obliquity, and perfect spin-orbit alignment; the latter assumption being observationally supported \citep{zhao2023measured} for the 55 Cancri star-planet system (in nomenclature: 55 Cancri A and e). The equations of \textit{InstellCa} have been modified to include longitude ($\delta$) in the updated \textit{InstellCa--2} code. The instellation at a given latitude ($\lambda$) on a planet is calculated by estimating the mean irradiance along the entire longitude on the 3D planet. Mathematically, it is expressed as,

\begin{equation}
    P(\lambda) = \frac{1}{2\pi} \int_{-\pi}^{\pi} I(\lambda, \delta) \, \mathrm{d}\delta
\end{equation}

\noindent The modified angle of incidence for an incident ray from any surface-element ($r_s$, $\theta$, $\phi$) on the star to a given point on the planet is then,
\begin{equation*}
    \cos{\alpha} = \vec{A}(\theta, \phi) \cdot \hat{r}_p
\end{equation*}

\noindent Or more simply,

\begin{equation} \label{new}
\cos\alpha = \frac{{d}^{\top}{n}}{\sqrt{{d}^{\top}{d}}}
\end{equation}

\noindent Where

\begin{equation}
    \rho = a - r_p \cos{\lambda} \cos{\delta}
\end{equation}

\begin{align}
    {d} &=
    \begin{pmatrix}
        r_s \cos\theta \cos \phi - \rho \\
        r_s \cos\theta \sin \phi - r_p \cos\lambda \sin\delta \\
        r_s \sin\theta - r_p \sin\lambda
    \end{pmatrix}, \\[6pt]
    {n} &=
    \begin{pmatrix}
        \cos\lambda \cos \delta \\
        \cos\lambda \sin \delta \\
        \sin\lambda
    \end{pmatrix}.
\end{align}

\noindent Equation \ref{new} is not a mere mathematical reformulation. It carries the physical equivalence of a cosine reduction factor \citep{berger1978long} as the hour angle of the star changes over the day.

\noindent If we assume a non-zero Bond albedo ($A_B$), the effective temperature as a function of latitude will be:

\begin{equation*}
    T(\lambda) = \left[ \frac{{P}(\lambda)(1 - A_B)}{\sigma} \right]^{1/4}
\end{equation*}

\noindent The mean flux  over the 3D spherical planet, $\overline{F}$, is computed by integrating ${P}(\lambda)$ with the weighting factor $\cos\lambda$:

\begin{equation}
    \overline{F} = \frac{\int_{-\pi/2}^{\pi/2} {P}(\lambda) \cos\lambda \, d\lambda}{\int_{-\pi/2}^{\pi/2} \cos\lambda \, d\lambda} = \frac{1}{2} \int_{-\pi/2}^{\pi/2} {P}(\lambda) \cos\lambda \, d\lambda
\end{equation}

\noindent Following the fast-rotator uniform redistribution framework \citep{cowan2011model}, the effective temperature is evaluated using the longitude-averaged flux $\overline{F}$:

\begin{equation*}
    \sigma T_{\mathrm{eff}}^4 = \overline{F}(1 - A_B)
\end{equation*}

\begin{equation}
    T_{\mathrm{eff}} = \left[ \frac{\overline{F}(1 - A_B)}{\sigma} \right]^{1/4}
\end{equation}

\section{Results}

\begin{figure}[h!]
    \centering
    \includegraphics[scale=0.4]{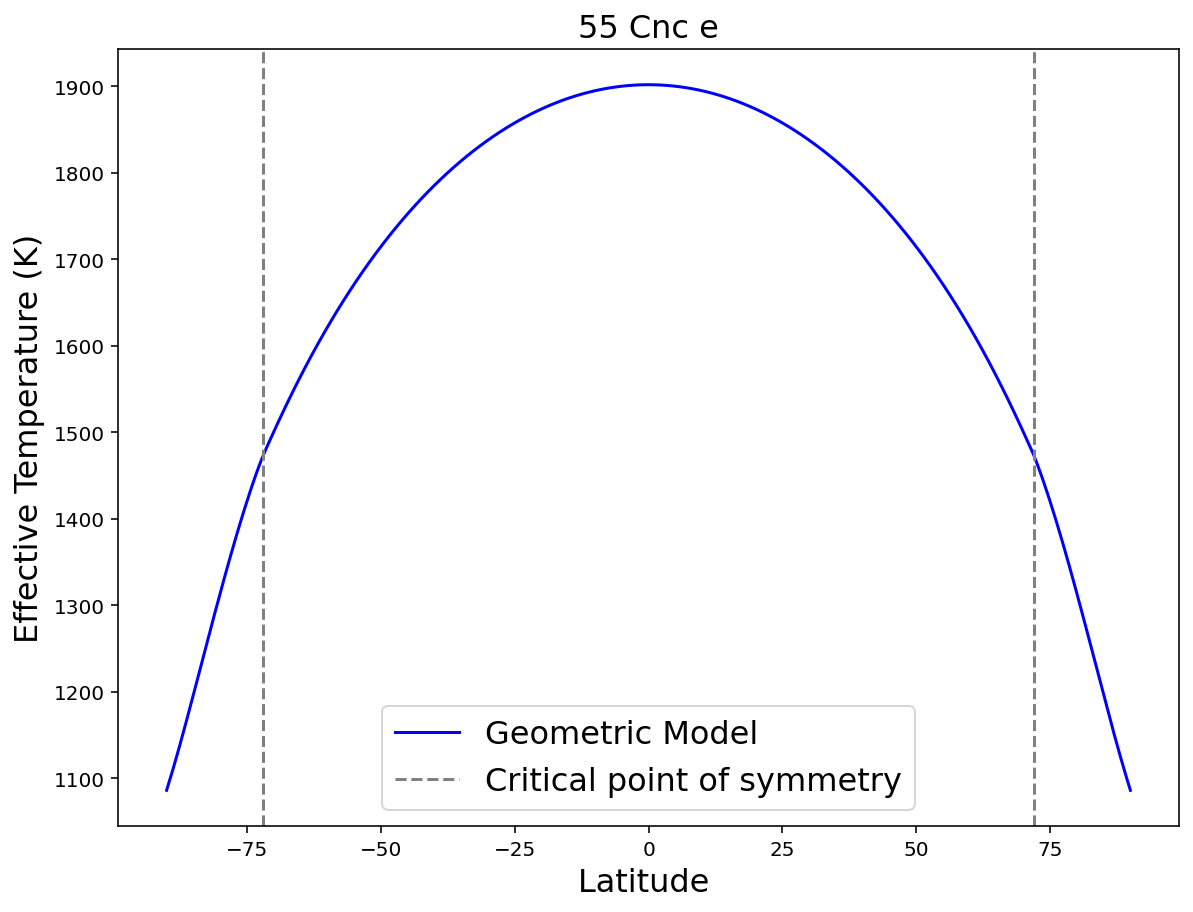}
    \caption{The geometry-modified thermal profile for 55 Cancri e with Bond albedo ($A_B=0.4$) assumption. The profile slope becomes less steep beyond the critical point of symmetry or in the penumbral zone. This is due to the fact that the penumbral zones always stay illuminated.}
    \label{example}
\end{figure}

\noindent \textit{InstellCa--2}\footnote{\href{https://github.com/Mradumay137/InstellCa\_2.0}{GitHub Repository}} provides longitude-averaged instellation values for all exoplanets in the current exoplanet archive \citep{schneider2011defining}. The resulting thermal profile for 55 Cancri e is shown in Fig. \ref{example}. In this scenario, it is observed that a starset never occurs in the penumbral regions.

Table \ref{one} shows the values of observationally estimated dayside temperatures from literature and the corresponding Bond albedos required to reproduce similar effective temperatures for the few close-in ultra-short period (USP) planets selected for this study.
If we assume that the dayside brightness temperature of 55 Cancri e, measured as $1796 \pm 88$ K, from the JWST MIRI spectra \citep{hu2024secondary}, to be the full-sphere effective temperature, we see that a rocky planet with a Bond albedo $A_B$=0.4 can naturally explain the assumed temperature. \citet{hu2024secondary} noted an 8$\sigma$ systematic uncertainty in their estimated dayside brightness temperature with the theoretical reference value of the effective temperature calculated under the assumption of inverse-square irradiance on a tidally locked bare rock. This inadvertently pushed realistic Bond albedos out of their modelling priors.
However, our geometric analysis shows that the systematic uncertainty can be reduced to 3$\sigma$ if the illumination reduces over the diurnal cycle with the zenith angle. In other words, the accurate reference value of the effective temperature is not $\sim$2770 K, as calculated through the inverse-square relation, but is rather $\sim$2200 K due to the rotational longitudinal redistribution after taking 3D geometry into account. The numerical error of the calculated effective temperatures in this work is negligible ($\sim$2--3 K) due to the analytic nature of the work. This error should not be confused with the statistical uncertainties of the stellar and planetary parameters adopted, which would influence the reference values of both the inverse-square and the 3D geometric model alike. The consequences of this geometric-rotational effect, therefore, also find application in modelling of systematic uncertainties that some other studies have performed \citep{fortune2025hot} for such planets.

\begin{figure*}[h!]
    \centering
    \includegraphics[width=2\columnwidth]{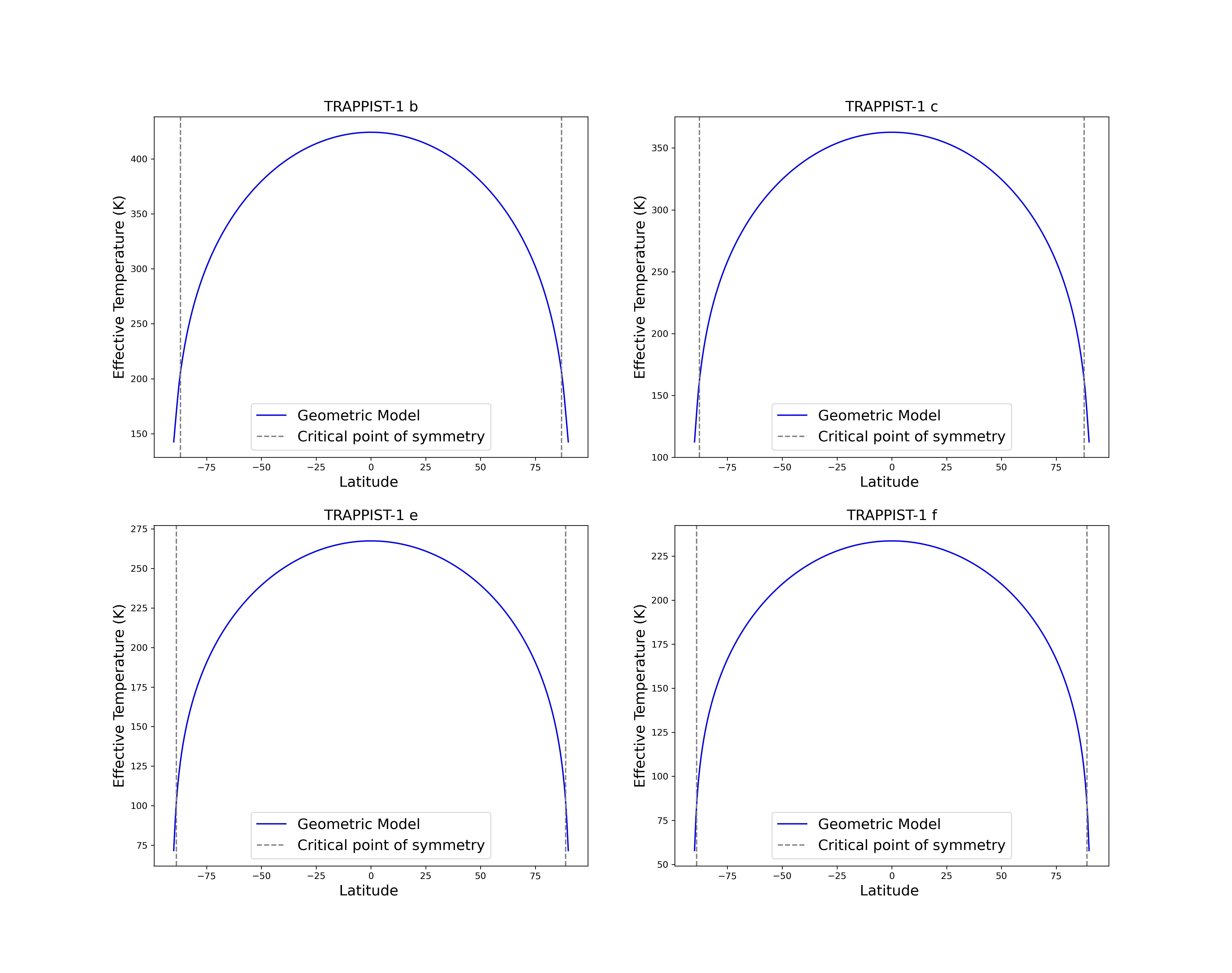}
    \caption{The geometry-modified thermal profiles for four TRAPPIST-1 system planets with the null Bond albedo assumption. The profiles align closely with the cosine assumption for far-off planets (TRAPPIST-1 e, TRAPPIST-1 f).}
    \label{three}
\end{figure*}

Interestingly, this means that the radiation geometry raises a degeneracy with the horizontal heat transport models of the planet. More importantly, this reinforces the idea that a greenhouse-warmed atmosphere \citep{angelo2017case,hammond2017linking} is not strictly necessary for explaining the detected temperature, which has also  been independently supported by geophysical and interior models for 55 Cancri e \citep{mercier2022revisiting,meier2023interior}. It must be noted here that this result does not eliminate the possibility of an atmosphere. On the contrary, the Bond albedo of 0.4 alludes to the expectations for terrestrial planet-like atmospheres \citep{rushby2019effect} that may not be necessarily participating in horizontal heat transport. It is also interesting to note that our model accurately reproduces the observed temperatures for the farther planets of the TRAPPIST-1 system (refer Table \ref{two} and Fig. \ref{three}), with the asynchronous rotation assumption and $A_B=0$, without invoking the need for the heat redistribution factor (f) that factors-in the value $(\frac{2}{3})^{1/4}$ \citep{burrows2008theoretical, koll2022scaling, lin2026persistent} for the no heat redistribution assumption on a tidally locked planet under point-source illumination.

\begin{table*}[h!]
\centering
\begin{tabular}{|l|c|c|c|}
\hline
Planet & Effective temperature $T_{eff}$ (K) & Literature estimate (K) & $A_B$ \\
\hline
K2-141 b  & 2069 & $2050 \pm 350$ & 0.20 \\
\hline
55 Cnc e  & 1794 & $1796 \pm 88$  & 0.40 \\
\hline
TOI-431 b & 1522 & $1520 \pm 350$ & 0.60 \\
\hline
TOI-2431 b & 2046 & $2063 \pm 30$  & 0.10 \\
\hline
TOI-561 b & 2136 & \begin{tabular}[c]{@{}c@{}}$1740 \pm 80$ (Eureka!)\\ $1830 \pm 70$ (JEDI 1)\\ $2150 \pm 80$ (JEDI 2)\end{tabular} & 0.30 \\
\hline
\end{tabular}
\caption{Effective temperatures, for a select few rocky candidates, derived from our model with Bond albedo kept as a free parameter to compare with literature values summarised in Table 1 of \citet{lin2026persistent}. The values are in turn derived from \citet{zieba2022k2,hu2024secondary, monaghan2025low,tacs2026earth, teske2025thick}.}
\label{one}
\end{table*}

\begin{table*}[htbp]
\centering
\begin{tabular}{|c|c|c|c|}
\hline
Planet & Effective temperature $T_{eff}$ (K) & Literature estimate (K) & Comparison Ratio ($\gamma$) \\
\hline
TRAPPIST-1 b & 399 & $470 \pm 17$  & 0.85 \\
\hline
TRAPPIST-1 c & 341 & $369 \pm 23$  & 0.92 \\
\hline
TRAPPIST-1 e & 251 & $251 \pm 4.9$ & 1    \\
\hline
TRAPPIST-1 f & 219 & $219 \pm 4.2$ & 1    \\
\hline
\end{tabular}
\caption{Effective temperatures under null Bond albedo ($A_B=0$), for the planets of the TRAPPIST-1 system, derived from our model and compared to literature values summarised in Table 1 of \citet{lin2026persistent}. Since the geometric correction reduces with decreasing angular size, we see that the values converge with the point-sized approximation and the ratio approaches the maximum value of unity, for the farther planets of the TRAPPIST-1 system, serving as a sanity check. The literature estimates are in turn derived from \citet{gillon2017seven, gillon2026no}.}
\label{two}
\end{table*}

\noindent We define a comparison ratio $\gamma$ as the ratio of effective temperature of the asynchronously rotating planet and the tidally-locked equilibrium temperature presented in observational studies \citep{gillon2017seven}.
It is evident that this comparison ratio ($\gamma$) shows degeneracy with the heat redistribution factor (f), as demonstrated with the exemplar TRAPPIST-1 system, under the assumption of null Bond albedo \citep{gillon2017seven}. In other words, one can, in principle, tweak \textit{f} or $\rm A_{B}$ to accommodate different atmospheric redistribution scenarios. However, the same can also be attributed to the combined effect of close-in geometry and rotational redistribution. It is worth re-iterating that the adoption of close-in geometry by itself implies considering a 3D spherical planetary surface, as opposed to a 2D disk that is quintessential in exoplanetary photometric research.

\section{Discussion}
\noindent The following part of the article focuses on the interpretation and analysis of results. Specifically, focusing on understanding them with the theoretical framework and benchmarking them with prior research. This also helps highlight the uniqueness of this work and how well it fits with the prior exoplanetary climate analysis models.  

\subsection{Theoretical Interpretation}

\begin{figure}[h!]
    \centering
    \includegraphics[width=\columnwidth]{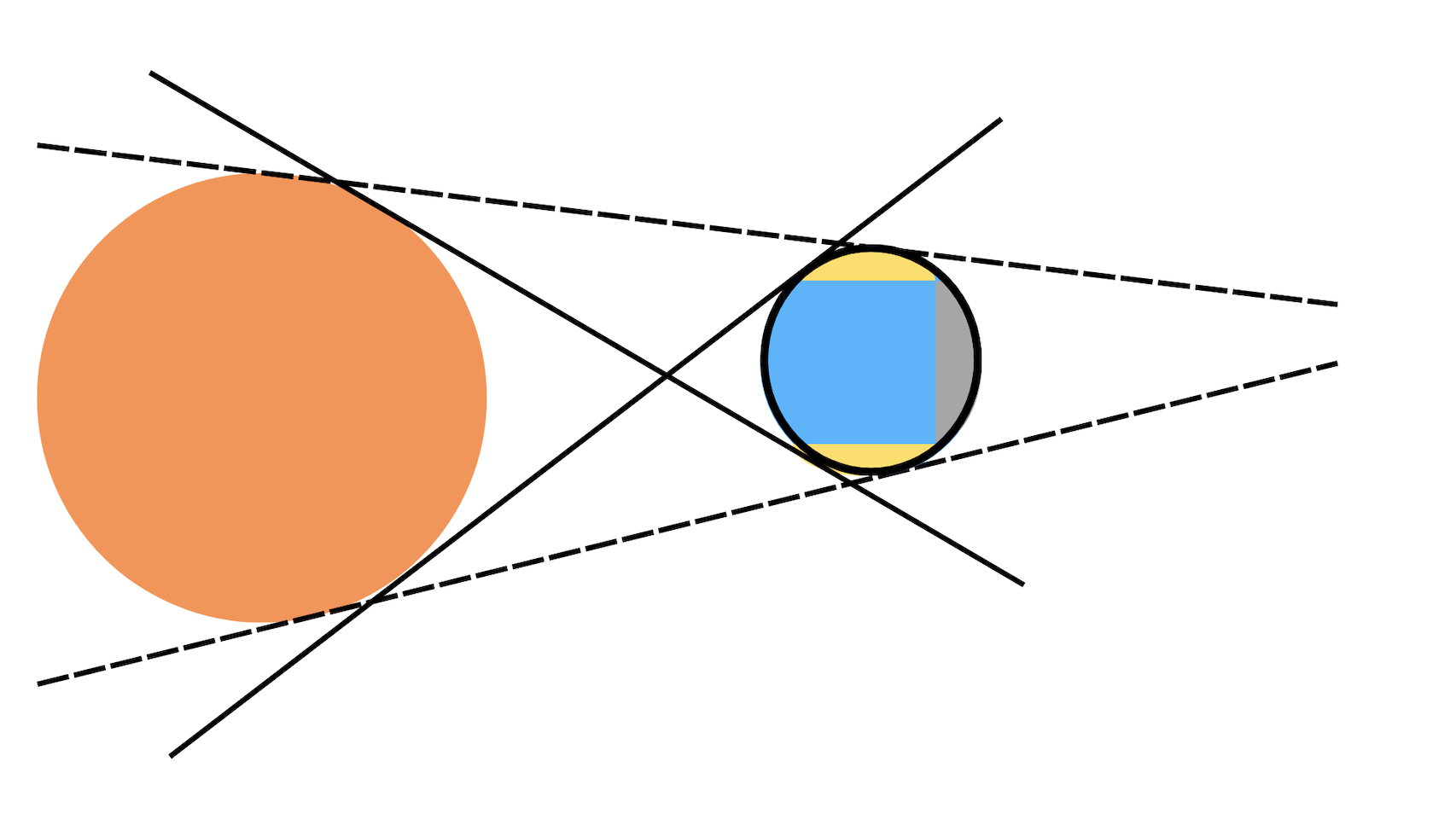}
    \caption{Schematic representation a star-planet system with the planet experiencing a day-night cycle including a \textit{starrise} and a \textit{starset}. The polar caps (indicated in yellow) remain illuminated throughout the cycle, hosting a steady-state temperature. The instantaneous night-side spherical cap (grey) is not permanent unlike the tidally-locked scenario. In other words, the penumbral regions are always illuminated and there is no permanent umbral region on the planet.}
    \label{main}
\end{figure}

\noindent To understand the case of ``forever-illuminated'' poles (depicted in Fig. \ref{main}) from a theoretical standpoint, with the framework developed by \citet{sadh2026importance}, we need to understand the radiation field geometry. As hypothesized by \citet{sadh2026importance}, the time-averaged Poynting vector (S) exhibits two components, namely radial and tangential.

\begin{equation}\label{tangential}
    \mathbf{S}= \alpha S_{r} \hat{r} + \beta S_{\theta} \hat{\theta},
\end{equation}

\begin{figure}[h!]
    \centering
    \includegraphics[width=\columnwidth]{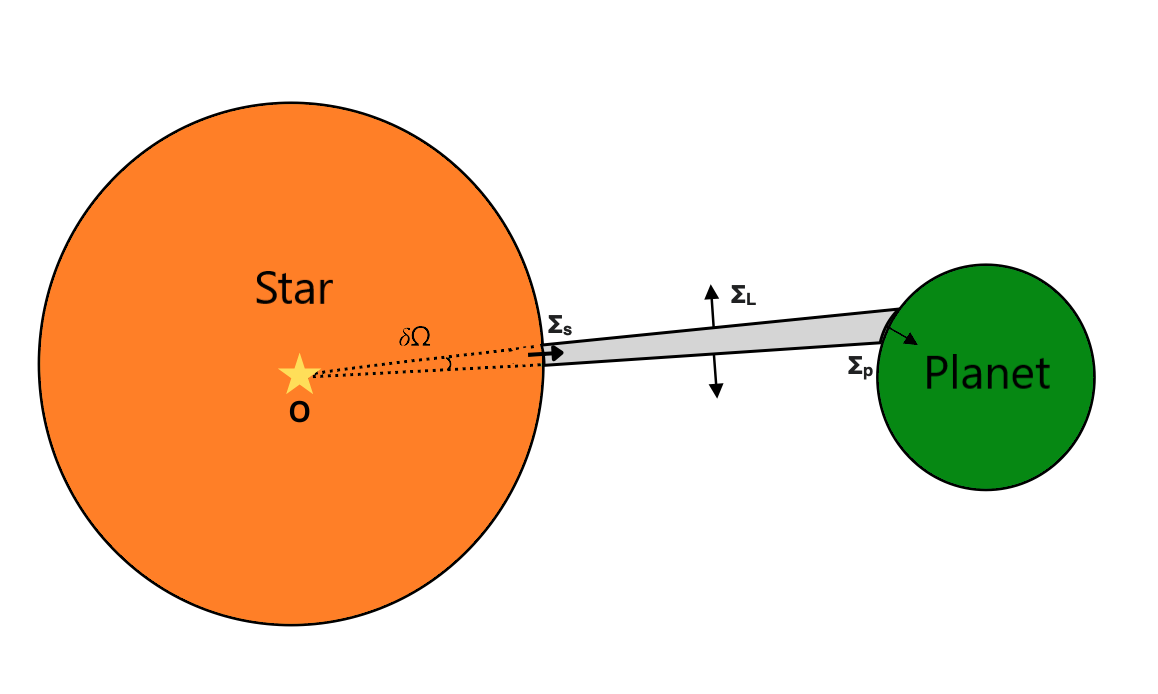}
    \caption{Originally adapted from \citet{sadh2026importance}. Schematic representation of the Gaussian 2D surface. An infinitesimal solid angle $\delta \Omega$ is centred in the centre of the star and reaches the surface of the planet. Along its path, it ``cuts out'' a portion of space between the star and the planet whose boundary is the surface under consideration. The arrows in the picture are the normal unit vectors to the surface. The yellow star denotes the star as equivalent to a point-sized source in case of spherical symmetry.}
    \label{gauss}
\end{figure}

\noindent Where $\alpha$ and $\beta$ are coefficients of each component, which depend on geometrical parameters and boundary conditions. In the penumbral zones of the rotating planet, the tangential component becomes the sole component contributing to the flux. In the penumbral zones, the radial flux crossing the Gaussian surface ($\Sigma_{p}$) (Fig. \ref{gauss}) is balanced out over a diurnal cycle. This can be understood by the following argument of symmetry. The hidden part of the star blocked beneath the local horizon at noon is compensated at dawn where the planet will see the same fraction of the star visible. The radiation field geometry is, therefore, equivalent to a plane-parallel field in polar angle space. This manifests with the radiative equilibrium temperature decreasing almost linearly at higher latitudes (Figures \ref{example}, \ref{three}). In other words, although the higher latitudes receive lower radiation flux, the size of the corresponding emitting belt on the planet scales down commensurately due to geometry. Therefore, the tangential component ($ \mathbf{S}_{\theta}$) proportionally balances the radiative emission by the penumbral cap on the planet. This results in a steady-state penumbral zone due to the thermal profile being flattened to near-linearity. Even in a tidally locked scenario, the polar temperature, albeit lower than this case, does not drop to zero, as demonstrated by \citet{sadh2026importance}. Therefore, this theoretical analogy applies to all cases where the point-source approximation fails and where the irradiance is approximated by Legendre polynomials \citep{kopal1954photometric, sadh2026importance}. In the far-field limit, the effective temperature approaches the point-source approximation value (refer Table \ref{two}).

\subsection{Thermal gradient in the penumbral zone}

\noindent Due to this flattening of the temperature slope in the penumbral zone (Figures \ref{example}, \ref{three}), the corresponding thermal gradient for a close-in planet approaches uniformity. Mathematically, this gradient can be defined as,

\begin{equation*}
    \nabla T_{\text{penumbra}} = \frac{\partial T}{\partial \lambda}
\end{equation*}

\noindent For close-in planets, an interesting consequence of a uniform gradient in the penumbral zone can be interpreted as follows. In addition to the atmospheric horizontal heat transport considerations being degenerate with geometry, the vertical heat transport from the planetary core to the crust, through geological and climate activity might also be significantly affected since the illuminated spherical polar caps host a steady-state temperate zone.

\subsection{Thermal Offset}

\begin{figure}[h!]
    \centering
    \includegraphics[width=1.05\columnwidth]{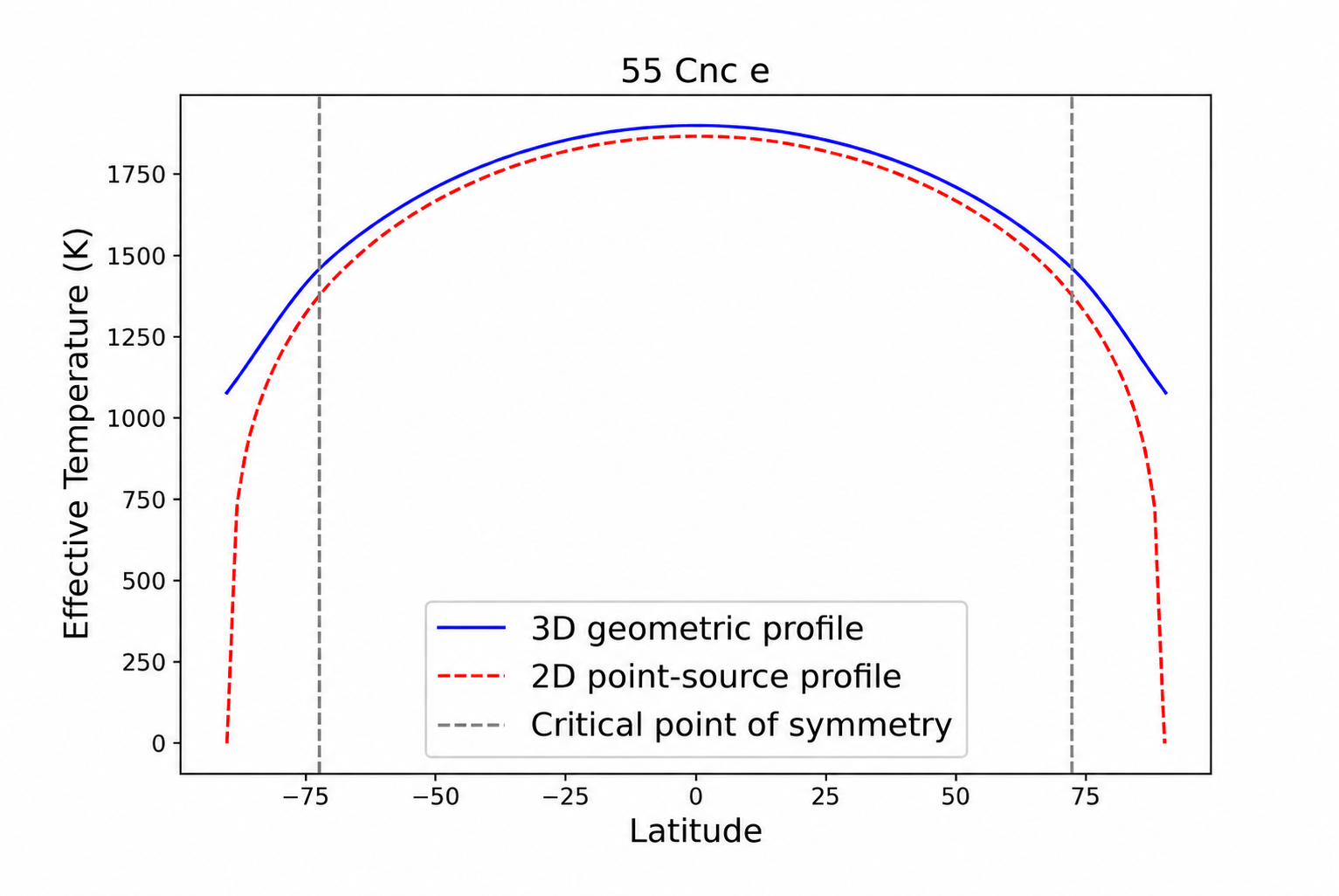}
    \caption{A clear thermal offset of 30.9 K is seen in the rotating close-in planet model. A phase curve model that only uses the point-source approximation underestimates the instellation on the planet. Interestingly, the illuminated twilight ring is the reason for this offset.}
    \label{baseline}
\end{figure}

\begin{figure}[]
\centering
\includegraphics[width=0.95\columnwidth]{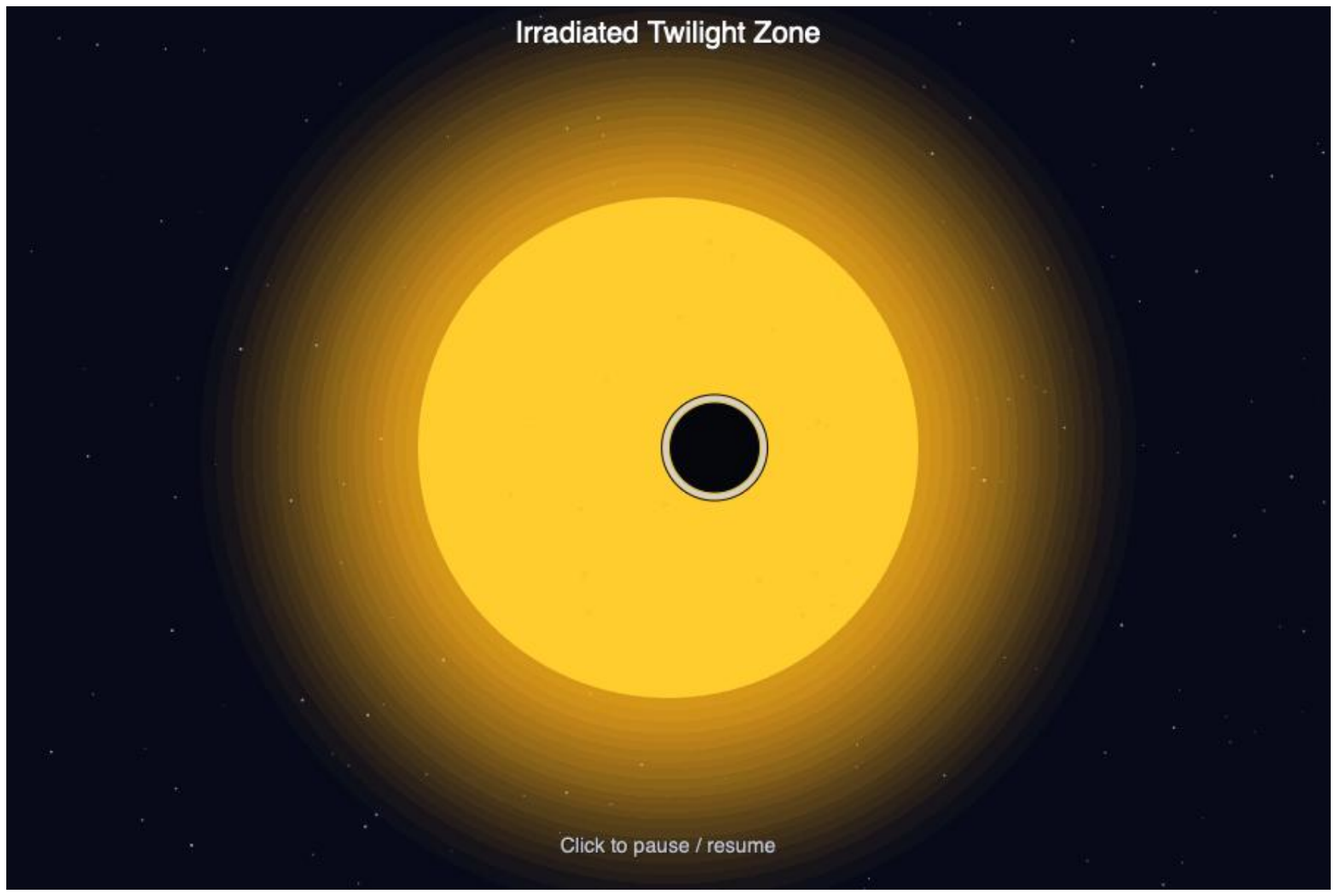}
\caption{Schematic representation of the penumbral ring effect from the animation snapshot. Due to the large angular size of the host star, the penumbral zone is most clearly seen between ingress and egress, creating a permanently irradiated twilight zone along the polar meridian without requiring atmospheric refraction. The animation was rendered using p5.js \citep{p5js}.}
\label{ring}
\end{figure}

\noindent The extended-source geometry leads to a pronounced twilight zone along the polar meridian of the planet. The prominence of this twilight zone changes how eclipses are conceived with an anthropocentric prior. In other words, instead of the planet being a completely dark disk occulting the star, it appears as a dark disk with a faintly illuminated penumbral ring. A snapshot (Fig. \ref{ring}) of the animated figure created to demonstrate this helps visualise how the eclipses on close-in planets show this \textit{penumbral ring-effect}.
Consequently, a thermal offset in the diurnal flux is observed for such close-in systems. Fig. \ref{baseline} shows the thermal offset for 55 Cancri e over one diurnal rotation period. The spin-orbit coupling of the planet will ultimately govern the net effect of this offset in the observed thermal phase curve.

The theoretical implication of this systematic flux offset might be geophysical in nature. For instance, a higher irradiation temperature at the sub-stellar point due to the contributions from the polar meridian penumbral ring may play a significant role for the modelled thermal phase offset \citep{meier2023interior}.

The solutions of the inverse-problem of determining the amplitude and phase offset of a thermal phase curve can be simplified if the degrees of freedom of the models are ascended with a 3D rotating planetary body. While the previous \textit{InstellCa} code helped with a benchmark for the day-night contrast amplitude, the inclusion of asynchronous rotation may explain the phase offset if spin-orbit resonance (2:1 or 3:2), as opposed to purely asynchronous rotation is assumed. In other words, the newly developed model allows for the irradiated flux differences to be prominent not only at the poles but also the sub-stellar point, which is coupled more strongly with the spin of the planet. However, a rigorous demonstration of this effect would only be possible if this model is integrated with the present phase curve models \citep{cowan2011model}.

\begin{figure}[h!]
    \centering
    \includegraphics[width=\columnwidth]{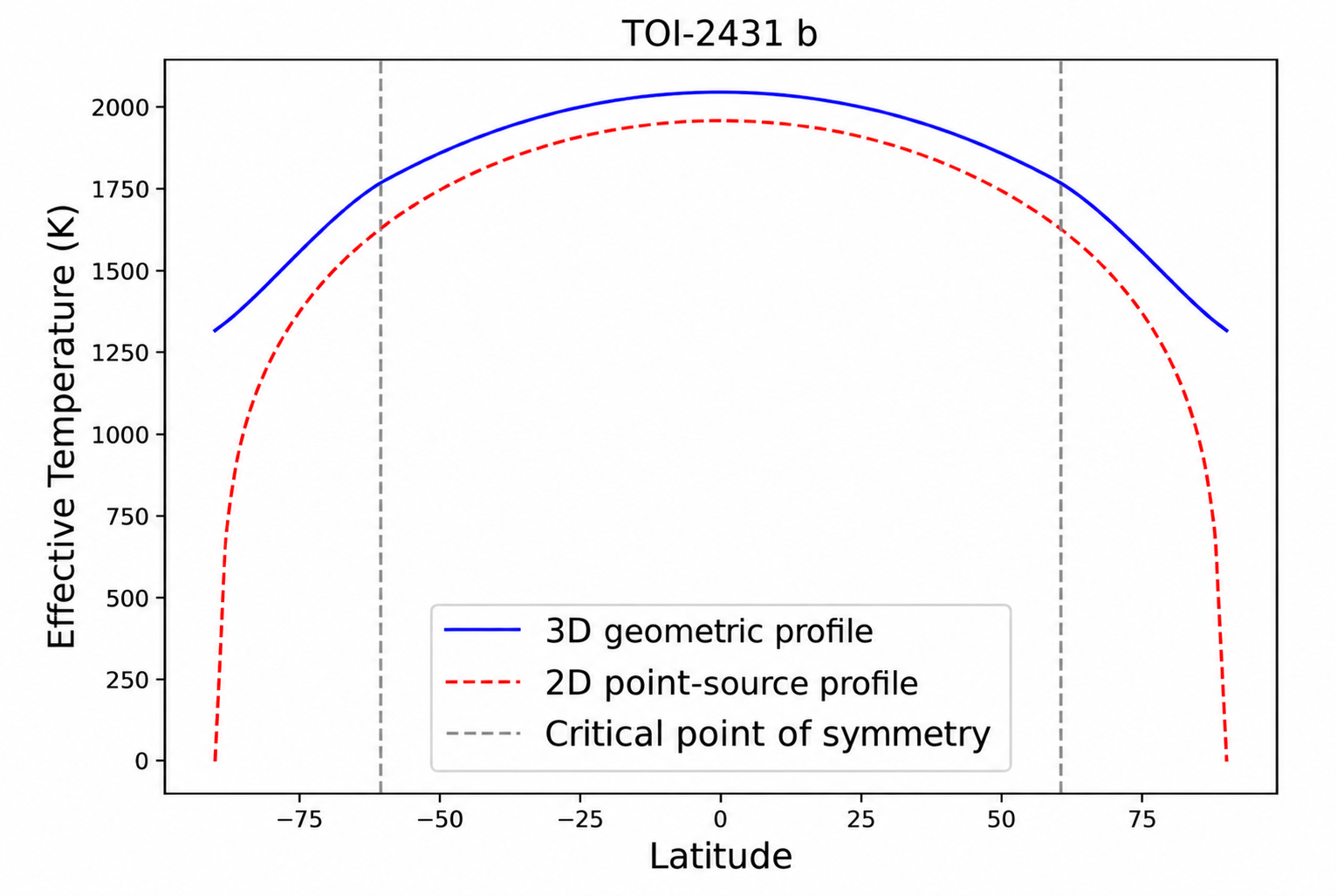}
    \caption{A pronounced offset (83.6 K) for the close-in planet TOI-2431 b. The terminator extends far beyond the poles, hyper-illuminating 67 percent of the planetary surface and indicating the most extreme case of the geometric effect.}
    \label{Extreme}
\end{figure}

\noindent The recently discovered planet TOI-2431 b \citep{tacs2026earth} helps to demonstrate the most extreme case of this thermal offset. Fig. \ref{Extreme} shows the instellation profile with a 70 K offset in the thermal flux at the sub-stellar point. The terminator for this planet is expected to extend approximately 118 degrees from the sub-stellar point. Our rotating planet model requires an albedo of 0.1 to reproduce the observed effective temperature \citep{tacs2026earth}.

\subsection{Reconciling the model with heat redistribution factor (f)}

\noindent Here, I demonstrate that the fast-rotator uniform redistribution case \citep{cowan2011model}, which is mathematically represented by f=1/4, is reproduced by the asynchronously rotating close-in planet model. 
\begin{equation}\label{former}
    \rm T^{*}_{\mathrm{eff}} = \left[ \frac{F_{IS}(1 - A_B)f}{\sigma} \right]^{1/4}
\end{equation}
Equation \ref{former} summarises the inclusion of the heat redistribution factor (f). Here $\rm F_{IS}$ is the flux according to the inverse-square approximation. \citet{burrows2008theoretical} developed the analytical foundations for the heat redistribution factor (f). However, the premise of their calculation relied on plane-parallel approximation of the incident rays. The geometrical model proposed in this work reconciles with that framework by modifying equation D13 of \citet{burrows2008theoretical} in altering the integral limits from ($sin[0]^{\circ},sin[90]^{\circ}$) to ($ sin[0]^{\circ}, sin[107]^{\circ}$) according to the extended day-night terminator. This gives the revised factor for uniform redistribution as f=0.27, which is only marginally higher than f=1/4. Using the revised redistribution factor we obtain the reference effective temperature for 55 Cancri e as:

\begin{equation*}
    \rm T^{*}_{\mathrm{eff}} = 1768 K
\end{equation*}

\noindent The difference from the $\rm T_{eff}$ value quoted in Table \ref{one} is attributed to the thermal offset (O), which is $\sim30 K$, as seen in Fig. \ref{baseline}. Mathematically this can be written as:

\begin{equation}
    \rm O=T_{eff}-T^{*}_{eff}
\end{equation}

\noindent Although this modification of the heat redistribution factor is modest, the emphasis of the back-of-the-envelope analysis lies in sanity checking the geometric model for consistency with pre-existing models.

\section{Conclusion}

This article serves to highlight the updated version of the code \textit{InstellCa} that now provides instellation derived temperature profiles for close-in planets, which has the potential to be integrated with the current 3D GCM models and used as an illustrative tool on exoplanet archives (e.g. Encyclopaedia of exoplanetary systems
\citep{schneider2011defining} and NASA exoplanet Archive \citep{akeson2013nasa}).

The success of the analysis lies in demonstrating that close-in hyper-illumination when coupled with asynchronous rotation raises a new aspect of non-zero sub-stellar thermal phase curve offset in addition to highlighting the effect on degeneracies with planetary heat transport mechanisms \citep{Teinturier2022}. Integration of this model with the present phase curve models will help reduce the corresponding atmospheric and geophysical modelling degeneracies for close-in planets like 55 Cancri e with debated climate states. The results have a strong theoretical basis rooted in radiative transfer geometry. The estimation of theoretically accurate instellation values and thermal profiles will be useful when being compared with high-precision estimations of effective temperatures derived from JWST observations of the present and future \citep{kempton2024transiting,hinkley2026exploration}.

\subsection{Breaking the Degeneracy: A Gedankenexperiment}

\noindent Let's imagine an observer standing on Earth. In this hypothetical scenario, the only tool this observer has is a detector that measures the magnitude of the time-averaged Poynting vector of incident radiation from the Sun. In physical terms, this detector would measure the insolation. The same insolation can be produced from different combinations of the distance, incidence angle, and intermediate absorption of the stellar radiation. So the observer can very well predict that a lower value of insolation on the detector might suggest the existence of clouds or atmospheric mechanisms that cause radiative attenuation. However, without further constraints, it would be difficult to predict whether a different geometry is at play, that is, either a distant star or a setting star far from the Zenith of the observer. Or, whether there is atmospheric extinction that reduces the magnitude of the Poynting vector. 

This elementary thought experiment sheds light on how a photometric observable exhibits an inherent degeneracy that has the potential to pervade into the systematic uncertainty of the associated method if one is not cautious to think from an alternate perspective to eliminate the bias. More importantly, this hints towards the potential role of polarisation in breaking the degeneracy. One can pursue a similar line of thought and ask; \textit{What characteristic polarisation would the interaction of the stellar radiation with the atmosphere produce?} The central constraint is that we can only measure the incident stellar radiation as a proxy for understanding the planetary radiative properties. This is where this idea discovers its limits. However, isolating infrared emission from the planet is highly feasible. Although the idea of characterising a planet's atmosphere through infrared polarimetry has been proposed in the past \citep{de2011characterizing}, this field requires more active thought and extensive research. However, the fingerprints of atmospheric scattering on the polarisation of radiation have been clearly well studied \citep{HansenTravis1974, HansenHovenier1974}.

\subsection{Future Work}
\noindent Future extensions of this work will investigate how the proposed planetary rotation can be coupled with the orbital phase. This will enable the \textit{InstellCa-2} code to be conducive to thermal phase curve modelling of close-in terrestrial planets.

\section*{Acknowledgments}

\noindent I acknowledge the inspiration from Ms. Trapti Atre in accomplishing this study. 

\section*{Code availability}

\noindent \textit{InstellCa-2} is made available as a web \href{https://instellca20-ubbrcrozf6fxct26rzqtez.streamlit.app}{application}.

\bibliographystyle{aa}
\bibliography{reference}

\end{document}